\documentclass[11pt]{article}
\usepackage{moriond,epsfig}

\def\be{\begin{equation}}
\def\ee{\end{equation}}
\def\bea{\begin{eqnarray}}
\def\eea{\end{eqnarray}}

\begin{document}
\vspace*{4cm}
\title{INCLUSIVE JET CROSS-SECTION MEASUREMENTS AT CDF}

\author{ O. NORNIELLA}

\address{Institut de F\'{i}sica d'Altes Energies, \\
Edifici Cn. Facultat Ci\`{e}ncies UAB, \\
E-08193 Bellaterra (Barcelona), SPAIN}

\maketitle\abstracts{
Results on inclusive jet production in proton-antiproton collisions at 
$\sqrt s $=1.96 TeV based on $1 fb^{-1}$ of CDF  Run II data are presented. 
Measurements are preformed using the $k_{T }$ algorithm in a wide range of jet transverse momentum 
and jet rapidity. The measured cross sections are compared to next-to-leading order perturbative QCD calculations.}

\noindent
The measurement of the inclusive jet cross section at the Tevatron is an important test of perturbative QCD (pQCD) 
predictions over more than 8 orders of magnitude, probing distances down to $10^{-19}$m. The increased 
center-of-mass energy in Run~II (from 1.8 TeV to 1.96 TeV), the highly upgraded CDF detector\cite{cdf}, 
and the amount of data collected allow jet measurements in an extended region of jet 
transverse momentum, $p_{T}^{jet}$, and jet rapidity, $y^{jet}$. Jet measurements at large rapidities are important because 
they are sensitive the gluon density in the proton in a kinematic region in $p_{T}^{\rm jet}$ where no effect from new physics is
expected.

\vspace{0.4cm}
This contribution presents results on inclusive jet production in five jet rapidity regions up to $|y^{jet}|$ = 2.1, 
based on $1 fb^{-1}$ of CDF Run II data. CDF used the longitudinally-invariant $k_{T}$ algorithm~\cite{kt} to search for jets:  
\begin{eqnarray}
k_i=p_{T,i}^{\rm 2}; \;\;\; \;\;\;k_{ij}=min(p_{T,i}^{\rm 2},p_{T,j}^{\rm 2})\cdot\frac{(y_i-y_j)^{\rm 2}+(\phi_i-\phi_j)^{\rm 2}}{D^{\rm 2}},
\end{eqnarray}
\noindent
where particles are clustered according to their relative transverse momentum. 
The algorithm includes a D parameter that approximately controls the size of the jet in the $\phi-y$ space.
These algorithm is infrared/collinear safe to all orders in pQCD and it does not need to 
solve situations with overlapping jets, making possible a better comparison between data and theory. 
A previous measurement using the $k_{T}$ algorithm at the Tevatron during Run I~\cite{D0RunI} observed 
a marginal agreement with NLO pQCD at low $p_{T}^{jet}$, thus suggesting the $k_{T}$ algorithm was particularly challenging 
in hadron collisions. However, these CDF results~\cite{kt_forward} show that this discrepancy is 
removed after non-perturbative corrections are included.  

\begin{figure}[htp]
\centerline{\includegraphics[scale=0.5,width=2.7in,height=2.8in]{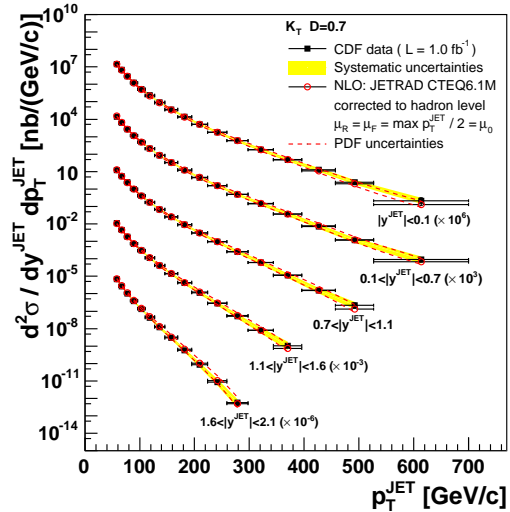}
}
\caption{Inclusive jet cross sections measured using the $k_T$ algorithm with D=0.7 for jets with $p_{T}^{jet}\geq 54 GeV/c$ 
in five rapidity regions up to  $| y^{jet}| = 2.1$. The black squares represent the measured cross sections and the shaded 
bands indicate the total systematic uncertainty on the data. The measurements are compared to NLO pQCD calculations. The 
dashed lines represent the PDFs uncertainties on the theoretical predictions}
\label{AllCSKt}
\end{figure}

\begin{figure}[p]
\centerline{\includegraphics[scale=0.5,width=3.1in,height=2.5in]{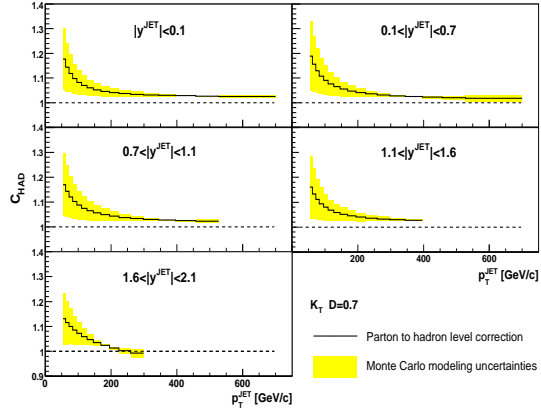}
}
\caption{Parton to hadron level corrections applied to the NLO calculations to correct for 
underlying event and hadronization contribution in the different rapidity regions. The shaded bands represents the associated 
uncertainty coming from the Monte Carlo modeling.}
\label{CHAFactors}
\end{figure}

\begin{figure}[p]
\vspace{-3cm}
\centerline{\includegraphics[scale=0.5,width=3.1in,height=2.5in]{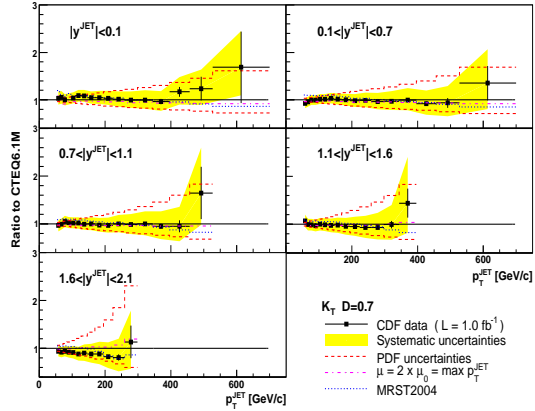}
}
\caption{Comparison between the measurements and the pQCD calculations. The dots are the ratios Data/Theory, 
the shaded bands indicate the total systematic uncertainty on the data and the dashed lines represent the PDFs uncertainties 
on the theoretical predictions.}
\label{RatioKt}
\end{figure}

\vspace{0.4cm}  
Figure~\ref{AllCSKt} shows the measured inclusive jet cross sections using the $k_T$ algorithm with D=0.7, for jets with 
$p_{T}^{jet}>$ 54 GeV/c in five jet rapidity regions up to $|y^{\rm jet}|=$ 2.1. For presentation, the different cross 
sections are scaled by a given factor. The measured cross sections have been corrected for detector effects back to the hadron 
level using PYTHIA-Tune~A Monte Carlo~\cite{pyt}, that provides an accurate description of the underlying event~\cite{pyta} and 
jet shapes~\cite{shapes} in Run II.  The cross sections decrease over more than seven orders of magnitude as $p_{T}^{jet}$ 
increases. The systematic uncertainties on the data, mainly dominated by a 2\%~to~3\% uncertainty in the jet energy scale, 
vary from 10\% at low $p_{T}^{jet}$ to about 50\% at high $p_{T}^{jet}$. The measurements are compared to NLO pQCD predictions 
as determined using JETRAD~\cite{jetrad} with CTEQ6.1M PDFs~\cite{pdf} and renormalization and factorization scales set to 
$p_{T}^{max}/2$, where $p_{T}^{max}$ is the $p_{T}$ of the leading jet. The theoretical calculations include 
correction factors, $C_{HAD}$, to take into account non-perturbative 
effects related to the underlying event and fragmentation processes. The factors, presented in figure~\ref{CHAFactors}, 
have been evaluated with PYTHIA-Tune~A as the ratios between the nominal cross sections at the hadron level and the 
ones obtained after turning off multiple parton interactions 
between remnants and fragmentation into hadrons. The difference obtained when HERWIG~\cite{hrw} is used instead of PYTHIA has 
been taken as the uncertainty on these factors. Figure~\ref{RatioKt} shows the ratios between the measurements 
and the theory are presented. A good agreement is observed over all $p_T^{jet}$ ranges in all rapidity regions. 
The uncertainty in the theoretical 
prediction is dominated by the uncertainty on the gluon PDF at high x which, at high $p_{T}^{jet}$, goes from $^{+70}_{-30}\%$ 
to $^{+140}_{-40}\%$ for central and forward jets, respectively. The uncertanties in the data compared to that 
in the NLO pQCD calculations show that the measurements will contribute to a better knowledge of the parton distributions inside 
the proton.

\vspace{0.6cm}

\clearpage

\begin{figure}[htp]
\vspace{-0.4cm}
\centerline{\includegraphics[scale=0.7,width=3.5in,height=2.85in]{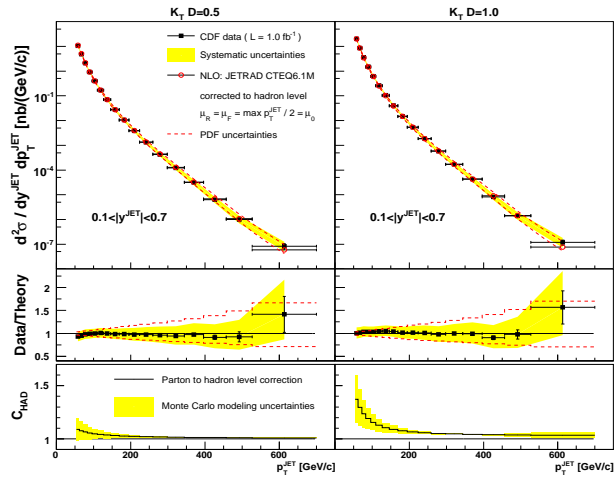}            
}
\caption{Inclusive jet cross sections measured using the $k_T$ algorithm with D=0.5 (left) and D=1.0 (right) 
for jets with $p_{T}^{jet}>$54 GeV/c and $0.1 <| y^{jet}|< 0.7$. The black squares represent the measured cross sections 
and the shaded bands indicate the total systematic uncertainty on the data. The measurements are 
compared to NLO pQCD calculations. The dashed lines represent the PDFs uncertainties on the 
theoretical predictions. The bottom plots show the parton to hadron level corrections 
applied to the NLO calculations to correct for underlying event and hadronization effects, where the shaded bands 
represent the associated uncertainty coming from the Monte Carlo modeling.}
\label{Central}
\end{figure}

For central jets, $0.1< | y^{jet}| < 0.7 $, the measurements are repeated using a D parameter equal to 0.5 and 1.0. As D increases, 
the average size of the jet in $\phi-y$ space increases, and the measurement becomes more sensitive to underlying event 
contributions. Figure~\ref{Central} shows the measurements. The good agreement still observed between the measured cross sections and 
the NLO pQCD predictions indicates that the soft contributions are well under control.

In summary, this contribution reports results on inclusive jet production in proton-antiproton collisions at $\sqrt s$ = 1.96 TeV, 
based on $1 fb^{-1}$ of CDF Run II data, using the $k_{T}$ algorithm. CDF also performed the measurement using the Midpoint cone-based 
algorithm~\cite{cone}. The measurements are in a good agreement 
with NLO pQCD calculations. In particular, for central jets and at high $p_{T}^{jet}$ no deviation with respect 
to the theory is found. In the most forward region, the total systematic uncertainty on the data is smaller than that 
on the theoretical calculations. Therefore, these new results will contribute to a better understanding of the gluon PDF 
in the proton at high x.


\section*{References}
\begin{thebibliography}{99}

\bibitem{cdf} R. Blair {\it et al.}, CDF Collaboration, 
{\it FERMILAB-Pub-96/390-E},
(1996).


\bibitem{kt} S. D. Ellis and D.E.Soper,
{\it Phys Rev. D},
{\bf 48}, 3160 (1993).


\bibitem{D0RunI} V. M. Abazov {\em et al.}, D0 Collaboration,
{\it Phys. Lett. B},
{\bf 525}, 211 (2002)   


\bibitem{kt_forward} A. Abulencia {\it et al.}, CDF Collaboration, 
Submitted to Phys. Rev. D, \\ 
hep-ex/0701051 (2007).


\bibitem{pyt} T. Sjostrand {\it et al},
{\it Comput. Phys. Commun.},
{\bf 135}, 238 (2001).

\bibitem{pyta}
R. D. Field,
{\it ME/MC Tuning Workshop.},
Fermilab, October 2002.

\bibitem{shapes}D. Acosta {\it et al}, CDF Collaboration,
{\it Phys Rev. D},
{\bf 71}, 112002 (2005).


\bibitem{jetrad} W. T. Giele {\it et al.},
{\it Phys Rev. Lett.},
{\bf 73}, 2019 (1994).

\bibitem{pdf}J. Pumplin {\it et al.},
{\it JHEP},
0207 (2002).

\bibitem{hrw}
G. Corcella {\it et al},
{\it JHEP},
{\bf 0101}, 010 (2001).


\bibitem{cone} A. Abulencia {\it et al}, CDF Collaboration,
{\it Phys Rev. D},
{\bf 74}, 071103(R) (2006).




\end{thebibliography}
\end{document}